%% file: shortpaper_revision_v3.tex
\documentclass[doublecol]{epl2} 
\usepackage{amsmath,amsfonts} 
\usepackage{graphicx,color} 

\let\a=\alpha \let\b=\beta

\let\s=\sigma  \let\f=\varphi \let\ph=\varphi
   \let\G=\Gamma
 \let\Th=\Theta  
   
 \let\r=\rho  \let\io=\infty

 \def\VV{{\cal V}}
\def\FF{{\cal F}}

\def\to{\rightarrow} \def\la{\left\langle} \def\ra{\right\rangle}

\newcommand{\beq}{\begin{equation}} \newcommand{\eeq}{\end{equation}}
\newcommand{\wh}{\widehat}

\title{A thermodynamic description of colloidal glasses}
\shorttitle{A thermodynamic description of colloidal glasses}

\author{Mauro Sellitto\inst{1} \and Francesco Zamponi \inst{2} }
\shortauthor{M. Sellitto \and F. Zamponi}

\institute{ \inst{1} Dipartimento di Ingegneria Industriale e
  dell'Informazione, Seconda Universit\`a di Napoli, Real Casa
  dell'Annunziata, I-81031 Aversa (CE), Italy \\ \inst{2} Laboratoire
  de Physique Th\'eorique, Ecole Normale Sup\'erieure, UMR 8549
  CNRS, 24 Rue Lhomond, 75005 Paris, France } 

\pacs{64.70.Q}{Theory and modeling of the glass transition}
\pacs{61.43.Fs}{Glasses} 
\pacs{82.70.Dd}{Colloids}

\abstract{ The phase behavior of hard-sphere particles interacting
  with a short-ranged potential is studied in the limit of infinite
  space dimensionality via the Franz-Parisi approach and the replica
  method of disordered systems. For an attractive square-well
  potential the phase diagram exhibits reentrancy of the liquid-glass
  transition, multiple glass states and glass-glass transition.  For a
  repulsive square shoulder potential no such special features are
  observed. Our results show that the Franz-Parisi approach can be
  consistently extended to deal with higher-order glass
  singularities and that interparticle attraction is crucial for
  complex glassy behavior in large enough dimensions, at
  least for monodisperse systems. }

\begin{document}

\maketitle

\section{Introduction}
Interacting hard spheres (with weak long-range attraction) have been
used since van der Waals to describe the gas-liquid phase
transition. More recently, they have been much studied in relation
with colloidal suspensions~\cite{Pusey}.  The interest has been especially
motivated by the possibility of introducing a (depletion-induced)
attraction between colloidal particles by adding a suitable amount of
non-adsorbing polymers to the suspension~\cite{Poon}. In doing so the
intensity and the range of attraction can be tuned by changing the
polymer concentration and the polymer coil radius, respectively. One
can thus explore a wide range of static and dynamic behaviors which
are not accessible in a single liquid system. When the range of
attraction is microscopic (i.e. of the order of a small fraction of
the sphere diameter) some fascinating large-scale macroscopic
properties emerge in such systems. They include the reversible
freezing-by-heating of the liquid phase and, at higher packing
density, structurally distinct types of dynamically arrested
states~\cite{Da00,Go09,ZP09,FSZT04,EB02,Ph02,Sc02,Frenkel}.  Such
features have been generally ascribed to structural changes of the
cage that confines particle motion. For weak attraction and high
packing density the glass formation is mainly driven by the usual
excluded-volume effects, while for strong attraction the tight
clustering of particles leads to an amorphous state with a rigid
gel-like structure which can exist even at very low packing. The
passage from one glass state to the other can occur either smoothly or
discontinuously (in the latter case the Debye-Waller factor undergoes
an extra jump).  Interestingly, similar complex glassy features have
been also identified in systems with no attractive 
interaction~\cite{SperlEma,Voigtmann,Ema,Se13}, where distinct glass states
with different packing properties arise solely from the competition
between crowding effects on different microscopic lenght scales.

In this paper we investigate the statistical mechanics of interacting
hard spheres in the case in which the space dimensionality is
infinitely large. In statistical physics high-dimensional spaces are,
in fact, a blessing. They make more feasible some
computations~\cite{Callen,Peliti} and, most importantly, provide a
limiting case in which the various mean-field approximations become
exact~\cite{Peliti,Parisi}. For this reason, the mean-field replica
approach we use is exact~\cite{KPZ12,PZ10} and this allows for a
comparison with the results obtained by alternative methods, such as
the Mode-Coupling Theory (MCT)~\cite{BF99,Fa99}, numerical
simulations~\cite{ZP09,FSZT04,Da00}, experiments~\cite{EB02,Ph02},
spin glass (or lattice glass)
models~\cite{CCN04,CL06,Kr07,CL07,KTZ08}, and the heterogeneous
facilitation picture~\cite{Se10,Se12,Se13}.  We generally find that
for short-range enough attraction the phase diagram exhibits
reentrancy of the liquid-glass transition line, multiple glass states
and glass-glass transition (much similar to those observed in
colloid-polymer mixtures, numerical simulations and MCT), while for
short-range repulsion no such special features is observed. Our
results show that the static approach based on the Franz-Parisi
potential can be consistently generalized to higher-order dynamical
glass singularities, and that attraction is an essential ingredient
for observing complex glassy features in high dimensional spaces.

\section{The model}

We consider a system of $d$-dimensional hard spherical particles of
unit diameter interacting with an attractive potential of constant
strength $U_0$ over a distance $\wh\s/d$: \beq v(r) = \begin{cases}
  \io & \text{for } r<1 \\ -U_0 & \text{for } 1<r<1+\wh\s/d \\ 0 &
  \text{for } r>1+\wh\s/d \\
\end{cases}
\eeq where $r$ is the inter-particle distance. Note that, in order to
have a non-trivial limit for $d\to\io$, the width of the attractive
part has been rescaled by the dimension according to the cage
size dependence on $d$ in the glass (i.e. proportionally to $1/d$),
see~\cite{PZ10} and Eq.~\eqref{eq:longMSD} below.

For a system of density $\r$ in the thermodynamic limit, we define
$\wh\f = 2^d \f/d$ where $\f = \frac{\pi^{d/2}}{\G(1+d/2)} \r$ is the
packing fraction of the repulsive core. The motivation behind this
scaling is that in the replica approach the glass transition occurs
for values of $\wh\f$ that remain finite for $d\to\io$ (e.g. in the
pure hard sphere case the dynamical transition happens at
$\wh\f=4.8$~\cite{PZ10}). We observe that MCT gives a different
scaling for the dynamical transition density, but numerical data seems
to support the replica result~\cite{CIPZ11}. Note, for comparison,
that it can be proven that sphere packings exist at least for $\wh\f
\leq 6/e$~\cite{Va11}, hence a $\wh\f$ of order 1 is a quite natural
scale for hard spheres in large dimensions (see~\cite{PZ10,TS10} for a
more detailed discussion).

Moreover, we fix $\b U_0 = 1/\wh T$ where $\b = 1/k_B T$.  The control
parameters are therefore rescaled density $\wh\f$, rescaled
temperature $\wh T$ and the width of the attraction $\wh\s$. In the
following we will also consider a ``sticky'' or ``Baxter'' limit in
which $\wh\s \to 0$ while the intensity of the attraction diverges,
hence $\wh T\to 0$, while $\mu = -1/\wh T - \log \wh\s$ is held
constant.

\section{Franz-Parisi potential}
\label{sec:II}

The basic idea of the replica approach to the glass transition is that
the kinetic slowing down on approaching the glass phase is due to the
sudden appearance of a bunch of long-lived metastable
states~\cite{KW87,KT88,KT89}. Under this assumption, the glass
transition can be identified by looking at the free energy of a {\it
  constrained} equilibrium system.  This is known as the Franz-Parisi
potential~\cite{FP95,FP97,CFP98}, see also~\cite{KT89,Mo95} for an
alternative but very related approach.  For particle systems, the best
way of computing this potential explicitly is the following. One
considers an equilibrium configuration $X = \{ x^\a_i \}_{i = 1, \cdots,
  N}^{\alpha=1, \cdots, d}$ of the liquid, and another configuration
$Y= \{ y^\a_i \}_{i = 1,\cdots, N}^{\alpha=1, \cdots, d}$ that is
constrained to be close to the first one, in such a way that the mean
square displacement is bounded by a fixed constant $\wh A$:
%
\beq\label{eq:longMSD} \frac{1}{2d} \frac1N \sum_{\alpha=1}^d \sum_{i=1}^N \la
(x^\a_i-y^\a_i)^2 \ra \leq \frac{\wh A}{d^2} \ . \eeq
One then computes the free energy of the system given $X$, and then
averages it over the equilibrium distribution of $X$. The result is
the average free energy $\VV_{\rm FP}(\wh A)$ of a system constrained
to be at distance $\wh A$ from a typical liquid configuration.  In the
liquid phase, particles diffuse away from any reference configuration,
and correspondingly $\VV_{\rm FP}(\wh A)$ has a unique minimum at $\wh
A = \io$. On the contrary, in the glass phase, particles are ``caged''
in a sphere of average radius $\sqrt{\wh A}$ around the reference
configuration and consequently there is a local minimum of $\VV_{\rm
  FP}(\wh A)$ at finite $\wh A$.  The {\it dynamical glass transition}
$T_{dyn}(\f)$ is signaled by the appearance of an inflection point in
$\VV_{\rm FP}(\wh A)$, and hence the secondary minimum appears
discontinuously at a finite $\wh A$ below $T_{dyn}(\f)$, and therefore
represents a generic {\it fold} singularity.  From an analytic point
of view, the computation of the Franz-Parisi potential requires the
use of the replica method~\cite{MP96,CFP98,MP09,PZ10}.  Since our
method is a direct extension of previous works~\cite{PZ10,BJZ11}, we
provide here only the final results of our computation.

The Franz-Parisi potential $\VV_{\rm FP}$ can be obtained
through the relation (more details of the derivation will be given
in~\cite{SeZa_Sapporo}):
\beq \label{VVder} \VV'_{\rm FP}(\wh A) = - \frac1{\wh A} \left[ 1 -
  \wh\f \FF_1(\wh A) \right] \ ,  \eeq
where
\beq\label{eq:F1phi} \FF_1(\wh A) = -\wh A \int_{-\io}^\io dy \, {\rm
  e}^{y} \frac{\partial q(\wh A;y)}{\partial \wh A} \log q(\wh A;y)
\eeq
and the function $q(\wh A;y)$ is 
\beq q(\wh A;y) = (1-e^{\wh U_0}) \Th\left(\frac{y+\wh A-\wh
  \s}{2\sqrt{\wh A}}\right) + e^{\wh U_0} \Th\left(\frac{y+\wh
  A}{2\sqrt{\wh A}}\right) \ . \eeq
The stationary points of $\VV_{\rm FP}(\wh A)$ are located at values
of $\wh A$ which are the solutions of $1/{\wh\f} = \FF_1(\wh A)$.  The
values of $\wh A$ at the local minima represent the long-time mean
square displacement (i.e. the Debye-Waller factor) in the glass.  In
this context the dynamical transition corresponds to the disappearance
of all the local minima of $\VV_{\rm FP}(\wh A)$, which happens when
$\wh\f^{-1} > \max_{\wh A} \FF_1(\wh A)$. Hence, the dynamic
transition is located at
\beq
\label{phid} 
\frac1{\wh\f_d} = \max_{\wh A} \FF_1(\wh A) \ , \eeq
and its stability requires the local (up) convexity of $\VV_{\rm FP}$.
A phase diagram obtained in this {\em static} framework is therefore
derived by studying the behavior of the functions $\VV_{\rm FP}$ and
$\FF_1$. Let us now consider two specific cases.

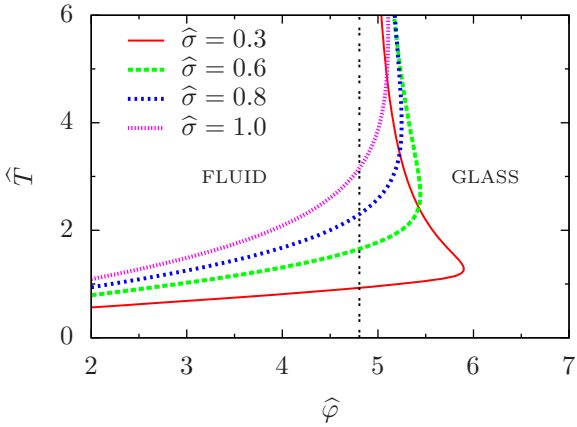
\begin{figure}
\input{reentrance_sw}
\caption{A section of the phase diagram in the rescaled variables,
  temperature $\wh{T}=k_{\scriptscriptstyle \rm B}T/U_0$ vs packing
  fraction $\wh\ph$, for attraction width $\wh\s$. The
  vertical line is the packing fraction at the dynamic glass
  transition for the purely hard-sphere potential. In this case there
  is only one glass phase but the fluid-glass transition line is
  reentrant: the liquid freezes upon heating.}
\label{phasediagram_sw}
\end{figure}

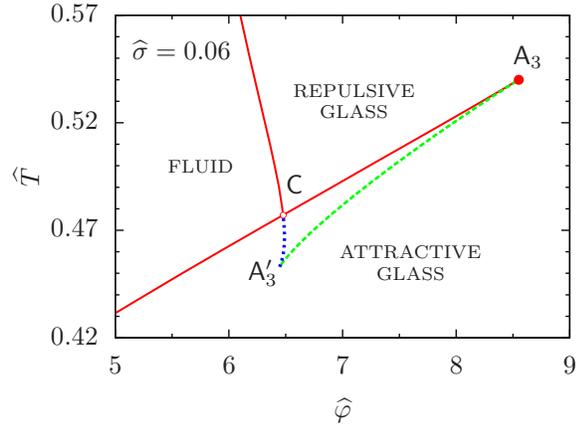
\begin{figure}[htbp]
\input{phase_sigma006}
\caption{Section of phase diagram at $\wh\s=0.06$.  There are
  two distint types of glass phases and fluid-glass transitions plus a
  glass-glass transition line.  The two fluid-glass transition lines
  meet at a crossing point ${\mathsf C}$ along with the glass-glass
  transition line.  The two special points ${\mathsf A}_3$ and
  ${\mathsf A}'_3$ are cusp singularities whose stability is
  determined by the Franz-Parisi potential.}
\label{fig:phase_sigma006}
\end{figure}

\section{Square-well potential}

We first discuss the results obtained for the square-well attractive
potential of width $\wh\s$ and depth $U_0$. We find that when the
attraction width is large enough (above $\wh\s \simeq 0.19$), the
function ${\cal F}_1$ has a single maximum for all densities and
temperatures, and the resulting phase structure is easily determined:
for each temperature, a single glass phase exists for densities larger
than $\wh\f_d$ defined by Eq.~\eqref{phid}.  

In Fig.~\ref{phasediagram_sw} we show the phase diagram in the
rescaled variables: reduced temperature, $\wh{T}=k_{\scriptscriptstyle
  \rm B}T/U_0$, and reduced packing fraction, $\wh\ph$. Interestingly,
for packing fractions above the dynamic glass transition of the purely
hard-sphere system the liquid-glass transition line develops a
reentrance: the liquid freezes upon heating. This reentrancy effect is
driven by the width of the square-well potential and is ultimately due
to entropic reasons. The smaller the attraction width the deeper the
fluid phase enters into the glass region.

\begin{figure}[h]
\input{FV_T047sigma006}
  \caption{Shape of the Franz-Parisi potential $\VV_{\rm FP}(\wh A)$ and
  the function $\FF_1(\wh A)$ in the region below $\wh{T}_{\mathsf
    C}$.  The
  three $\VV_{\rm FP}(\wh A)$ curves correspond to the points at
  which the phase diagram transition lines intersect the horizontal
  segment at $\wh T=0.47$. The dots denote the stationary points of
  the function $\FF_1(\wh A)$ and the corresponding ones in $\VV_{\rm
    FP}(\wh A)$.  Notice that $\FF_1$ does not depend on density while
  $\VV_{\rm FP}$ does.}
\label{fig:FV_sigma006}
\end{figure}
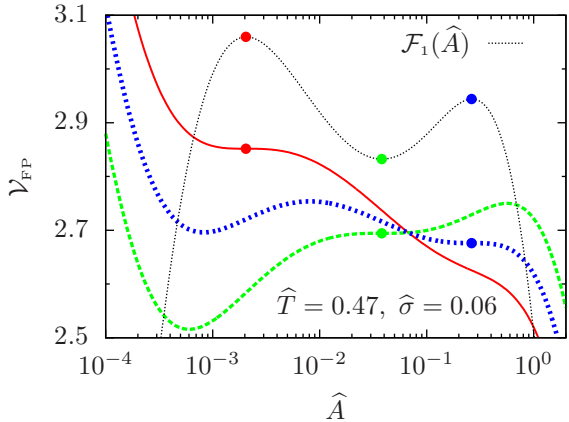

When the attraction width is below $\wh\s \simeq 0.19$, the function
${\cal F}_1$ shows two maxima in a range of temperatures and
densities, and the determination of the phase diagram requires some
care because of the appearance of multiple glass phases.  For
simplicity we focus on a representative case, obtained by fixing
$\wh\s = 0.06$. The corresponding section of the phase diagram is
shown in Fig.~\ref{fig:phase_sigma006}. There are two lines, ${\mathsf
  A}_3$-${\mathsf C}$ and ${\mathsf A}'_3$-${\mathsf C}$ separating
the fluid phase from two distinct types of glass phase, named {\it
  attractive glass} (that with smaller $\wh A$) and {\it repulsive
  glass} (the one with larger $\wh A$). These lines, which correspond
to the maxima of ${\cal F}_1$, cross each other at the point ${\mathsf
  C}$ and terminate at the critical endpoints ${\mathsf A}_3$ and
${\mathsf A}'_3$. The curve ${\mathsf A}_3$-${\mathsf A}'_3$ instead
corresponds to the minima of ${\cal F}_1$. The closed area delimited
by the lines going through the points ${\mathsf A}_3$, ${\mathsf C}$
and ${\mathsf A}'_3$ represents the coexistence region of the two
glass phases, while outside this region only one glass is present.  It
is important to observe that the ${\mathsf A}'_3$ point corresponds to
a {\it local quartic maximum} of the Franz-Parisi potential: it is
therefore dynamically unobservable as the repulsive glass will always
be unstable around this singularity.  On the contrary, the ${\mathsf
  A}_3$ singularity corresponds to a {\it local quartic minimum} of
the Franz-Parisi potential, and is therefore stable.  The detailed
shape of $\VV_{\rm FP}$ in the former case is shown in
Fig.~\ref{fig:FV_sigma006} for densities located at the intersection
of the transition lines with the horizontal segment $\wh T = 0.47$.
We notice that our conclusion is consistent with the results of MCT
and, interestingly enough, that the MCT dynamic stability criterion is
automatically built up in the Franz-Parisi potential: indeed, the
minima of the Franz-Parisi potential correspond to dynamically stable
solutions in MCT.  This static approach can be therefore consistently
extended to the determination of higher-order dynamic glass
singularities.

\begin{figure}[t]
\input{OP_sigma006T047}
\caption{Cage radius $\sqrt{\wh{A}}$ vs packing density $\wh\ph$ for
  attraction width $\wh\s=0.06$ and temperatures slightly below and
  above the point $\mathsf C$.  
}
\label{fig:OP}
\end{figure}
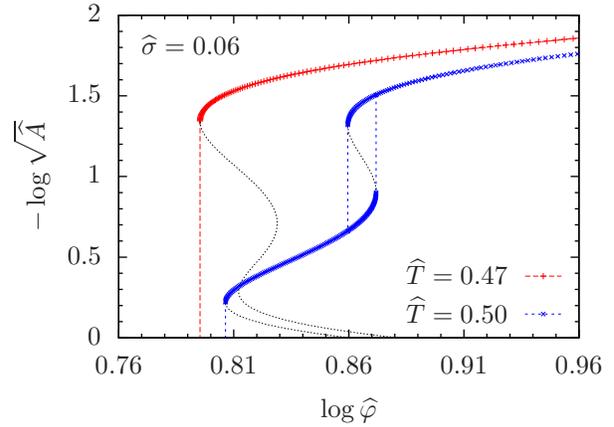

For completeness we also show the related evolution of the cage radius
$\sqrt{\wh{A}}$ as a function of the packing density in
Fig.~\ref{fig:OP} at temperatures slightly above and below the
${\mathsf C}$ point.  The double jump observed at $\wh{T}=0.5$
represents the transition from the liquid to the repulsive glass
followed, for increasing packing density, by the one from the
repulsive glass to the attractive glass.  As it happens with the usual
first-order transition hysteresis effects between the two glass phases
should be observed. At $\wh{T}=0.47$ there is only the
liquid-to-attractive glass transition.  In this latter case if by any
chance the appearance of the attractive glass is delayed, the liquid
could momentarily transform into a repulsive glass.

\begin{figure}[t]
\input{phase_global}
\caption{Global phase diagram in the two-glass region for several
  values of $\wh\s$. The sticky limit is represented by a full line.
  The two dotted light lines departing from the swallowtail
  bifurcation ${\mathsf A}_4$ are the line of cusp singularities
  ${\mathsf A}_3$ (represented as full dots) the line of crossing
  points ${\mathsf C}$ (empty dots). }
\label{phase_global}
\end{figure}
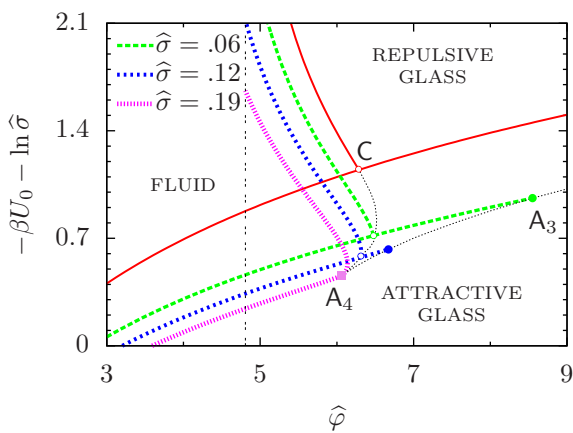

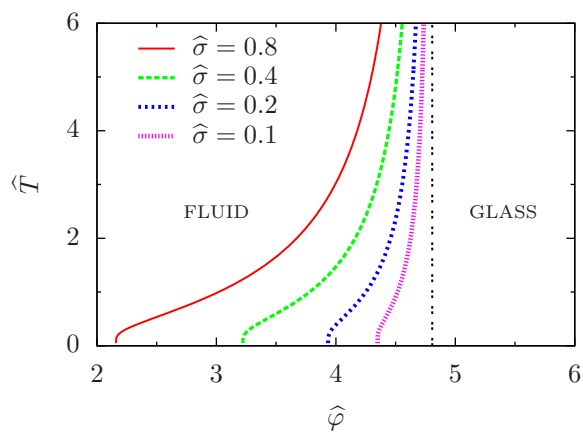
\begin{figure}[t] 
\input{phasediagram_ss}
\caption{Phase diagram for the square-shoulder repulsive potential for
  several values of the width $\wh\s$.}
\label{fig:phase_ss}
\end{figure}

Fig.~\ref{phase_global} summarizes the global structure of the phase
diagram in the two-glass region.  Upon decreasing $\wh\s$ the
glass-glass transition line shortens and eventually disappears at
$\wh\s \approx 0.19$ where the ${\mathsf A}_3$ cusp merges into a
singularity of type ${\mathsf A}_4$, also known as swallowtail
bifurcation.  From Eq.~\eqref{VVder} it is easy to see that at the
stationary points of the potential, the vanishing of the $\ell$-th
derivative of $ \FF_1(\wh A)$ leads to the vanishing of the
$(\ell+1)$-th derivative of $\VV_{\rm FP}(\wh A)$.  The higher-order
singularities, ${\mathsf A}_3$ and ${\mathsf A}_4$, are special
singular points of the Franz-Parisi potential. Near these higher-order
singularities peculiar slow logarithmic relaxation occurs, a behavior
that has been explored in great detail in the MCT
framework~\cite{Da00,Go09} and
numerically~\cite{CR07,STZ03,ZP09,FSZT04,Da00}.  
  Fig.~\ref{phase_global} also includes the Baxter (or sticky sphere)
  limit: $\wh\s \to 0, \, U_0 \to \infty$ (or equivalently $\wh T \to
  0$) while $\mu = -\b U_0 - \log \wh\s = -1/\wh T - \log \wh \s$ is
  kept constant.   In this limit the small-$\wh A$ maximum of
$\FF_1(\wh A)$, corresponding to the attractive glass, moves to $\wh
A=0$, meaning that particles are completely frozen. The height of this
maximum can be computed explicitly and gives $\wh\f_{\rm d} = 2
e^{\mu}$.  Hence in this case the cusp singularity ${\mathsf A}_3$
moves to infinite packing density. This implies that there is no
smooth path in the phase diagram from the repulsive to the attractive
glass.  The resulting curve is reported in Fig.~\ref{phase_global} as
a full line.  Notice that in the limit $d\to \io$ the Kauzmann
transition is pushed at infinite packing density and is therefore
outside the range of physical relevance. Very similar phase diagrams
have been first obtained by MCT~\cite{Da00,Go09}, and later confirmed
by numerical simulations~\cite{ZP09,FSZT04} and
experiments~\cite{EB02,Ph02}.  In a previous attempt to
  use the replica method for this problem, based on the replicated
  Hypernetted Chain approximation and in $d=3$, no glass-glass
  transition was found~\cite{VPR06}, arguably due to the rather poor
  accuracy of the replicated Hypernetted Chain approximation in the
  phase region where the ``cage size'' is very small~\cite{PZ10}.

\section{Square-shoulder potential}

The analysis carried out above for the square-well potential can be
immediately extended to the square-shoulder potential by exchanging
$U_0 \to -U_0$. This case is interesting because it provides the first
finite dimensional ($d=3$) instance in which multiple glasses are
observed with only purely repulsive potential~\cite{SperlEma}. Similar
features have been recently obtained also in some binary
mixtures~\cite{Voigtmann} and in the heterogeneous facilitation
approach~\cite{Se13}.  The phase diagram for the square-shoulder
potential in infinite dimension is reported in
Fig.~\ref{fig:phase_ss}. Perhaps surprisingly, we find no trace of
distinct glasses but only the obvious widening of the glass phase when
the repulsion gets stronger (or the temperature is lowered)
and the interaction range increases.

  This negative result suggests that the contribution of ``ring
  diagrams'', which vanish in the limit of high
  dimensionality~\cite{FP99,PZ10}, is relevant for the appearance of
  multiple glasses in purely repulsive potential in finite dimension
  $d$.  Indeed, in Ref.~\cite{SperlEma} it is argued that the complex
  phase diagram of the square-shoulder potential is intimately
  connected to the competition between the amplitudes of the jumps of
  the pair correlation $g(r)$ at the two contact points corresponding
  to the two singularities of the potential (see Fig.~3 and the
  associated discussion in Ref.~\cite{SperlEma}).  This effect, that
  is clearly observed in $d=3$, disappears in infinite dimensions
  where all ring diagrams vanish: in this case, in the liquid phase,
  one has $g(r) = \exp[-\b v(r)]$, the structure of $g(r)$ is much
  simpler, and in particular the value of $g(r)$ at the first contact
  decreases on increasing the attraction strength and is always lower
  than the one at the second contact.  Clearly, the richer structure
  of $g(r)$ in $d=3$, which is responsible for the complex phase
  diagram of the square-shoulder potential, is due to the contribution
  of ring diagrams (i.e., many-body effects) which are absent in
  infinite dimensions.

  This leads us to conclude that, while the complex phase diagram of
  the square-well attractive potential is well accounted for by the
  simple two-body mean field approximation in infinite dimensions,
  capturing the same physics for the square-shoulder potential
  requires the inclusion of (static) many-body effects that are only
  present in low enough dimensions~\cite{SperlEma}.  These are
  automatically implemented in MCT through the static structure factor
  of the liquid which is taken as input of the theory; the same can be
  done in principle in the replica approach but it requires a
  systematic improvement of the simplest approach used in this paper,
  along the lines discussed in~\cite{PZ10}.

\section{Conclusion}

The main result of this paper is that a static replica picture based
on the Franz-Parisi potential allows to re-derive many of the results
that have been previously obtained using MCT for attractive
colloids~\cite{Da00,Sc02,Go09}, namely the reentrance of the glass
transition line and two distinct glass phases for very short range
attractions. Here we limited ourselves to the $d\to\io$ limit where
computations are easier and the replica theory is exact~\cite{KPZ12},
at least according to the standards of theoretical physics.  For soft
matter applications, one would like of course to extend this
calculation in $d=3$~\cite{PZ10}. We expect that the phase diagram
will remain qualitatively the same, at least for the square-well
attractive potential. 
For the square-shoulder potential,
  we expect that including properly the finite dimensional
  contributions should allow one to recover the complex phenomenology
  discovered in~\cite{SperlEma}.  Unfortunately, constructing a good
  approximation scheme for the dynamical transition in low dimensions
  within the small cage expansion scheme is a non-trivial task, but
  there is hope to achieve it and work is in progress in this
  direction. Moreover, one could compute within this framework the
equation of state of the two glasses, the jump of specific heat at the
glass transition, and address the role of the Kauzmann transition.

\acknowledgments

This work originated in Ein Gedi during the Italo-Israelian bilateral
meeting on {\it Statistical Physics of Glass Formation and Amorphous
  Solids}.  We thank R. Benzi, G. Parisi and I. Procaccia for that
opportunity. We also thank P.~Charbonneau and E.~Zaccarelli for many
useful discussions.

\bibliographystyle{eplbib}

\end{document}

%% file: reentrance_sw.tex
\begingroup
  \makeatletter
  \providecommand\color[2][]{%
    \GenericError{(gnuplot) \space\space\space\@spaces}{%
      Package color not loaded in conjunction with
      terminal option `colourtext'%
    }{See the gnuplot documentation for explanation.%
    }{Either use 'blacktext' in gnuplot or load the package
      color.sty in LaTeX.}%
    \renewcommand\color[2][]{}%
  }%
  \providecommand\includegraphics[2][]{%
    \GenericError{(gnuplot) \space\space\space\@spaces}{%
      Package graphicx or graphics not loaded%
    }{See the gnuplot documentation for explanation.%
    }{The gnuplot epslatex terminal needs graphicx.sty or graphics.sty.}%
    \renewcommand\includegraphics[2][]{}%
  }%
  \providecommand\rotatebox[2]{#2}%
  \@ifundefined{ifGPcolor}{%
    \newif\ifGPcolor
    \GPcolortrue
  }{}%
  \@ifundefined{ifGPblacktext}{%
    \newif\ifGPblacktext
    \GPblacktexttrue
  }{}%
  \let\gplgaddtomacro\g@addto@macro
  \gdef\gplbacktext{}%
  \gdef\gplfronttext{}%
  \makeatother
  \ifGPblacktext
    \def\colorrgb#1{}%
    \def\colorgray#1{}%
  \else
    \ifGPcolor
      \def\colorrgb#1{\color[rgb]{#1}}%
      \def\colorgray#1{\color[gray]{#1}}%
      \expandafter\def\csname LTw\endcsname{\color{white}}%
      \expandafter\def\csname LTb\endcsname{\color{black}}%
      \expandafter\def\csname LTa\endcsname{\color{black}}%
      \expandafter\def\csname LT0\endcsname{\color[rgb]{1,0,0}}%
      \expandafter\def\csname LT1\endcsname{\color[rgb]{0,1,0}}%
      \expandafter\def\csname LT2\endcsname{\color[rgb]{0,0,1}}%
      \expandafter\def\csname LT3\endcsname{\color[rgb]{1,0,1}}%
      \expandafter\def\csname LT4\endcsname{\color[rgb]{0,1,1}}%
      \expandafter\def\csname LT5\endcsname{\color[rgb]{1,1,0}}%
      \expandafter\def\csname LT6\endcsname{\color[rgb]{0,0,0}}%
      \expandafter\def\csname LT7\endcsname{\color[rgb]{1,0.3,0}}%
      \expandafter\def\csname LT8\endcsname{\color[rgb]{0.5,0.5,0.5}}%
    \else
      \def\colorrgb#1{\color{black}}%
      \def\colorgray#1{\color[gray]{#1}}%
      \expandafter\def\csname LTw\endcsname{\color{white}}%
      \expandafter\def\csname LTb\endcsname{\color{black}}%
      \expandafter\def\csname LTa\endcsname{\color{black}}%
      \expandafter\def\csname LT0\endcsname{\color{black}}%
      \expandafter\def\csname LT1\endcsname{\color{black}}%
      \expandafter\def\csname LT2\endcsname{\color{black}}%
      \expandafter\def\csname LT3\endcsname{\color{black}}%
      \expandafter\def\csname LT4\endcsname{\color{black}}%
      \expandafter\def\csname LT5\endcsname{\color{black}}%
      \expandafter\def\csname LT6\endcsname{\color{black}}%
      \expandafter\def\csname LT7\endcsname{\color{black}}%
      \expandafter\def\csname LT8\endcsname{\color{black}}%
    \fi
  \fi
  \setlength{\unitlength}{0.0500bp}%
  \begin{picture}(4680.00,3402.00)%
    \gplgaddtomacro\gplbacktext{%
      \csname LTb\endcsname%
      \put(550,704){\makebox(0,0)[r]{\strut{}$0$}}%
      \put(550,1515){\makebox(0,0)[r]{\strut{}$2$}}%
      \put(550,2326){\makebox(0,0)[r]{\strut{}$4$}}%
      \put(550,3137){\makebox(0,0)[r]{\strut{}$6$}}%
      \put(682,484){\makebox(0,0){\strut{}$2$}}%
      \put(1402,484){\makebox(0,0){\strut{}$3$}}%
      \put(2122,484){\makebox(0,0){\strut{}$4$}}%
      \put(2843,484){\makebox(0,0){\strut{}$5$}}%
      \put(3563,484){\makebox(0,0){\strut{}$6$}}%
      \put(4283,484){\makebox(0,0){\strut{}$7$}}%
      \put(176,1920){\rotatebox{-270}{\makebox(0,0){\strut{}$\wh{T}$}}}%
      \put(2482,154){\makebox(0,0){\strut{}$\wh{\varphi}$}}%
      \put(1762,1921){\makebox(0,0){\scriptsize FLUID}}%
      \put(3653,1921){\makebox(0,0){\scriptsize GLASS}}%
    }%
    \gplgaddtomacro\gplfronttext{%
      \csname LTb\endcsname%
      \put(1364,2946){\makebox(0,0)[l]{\strut{}$\widehat{\sigma}=0.3$}}%
      \csname LTb\endcsname%
      \put(1364,2726){\makebox(0,0)[l]{\strut{}$\widehat{\sigma}=0.6$}}%
      \csname LTb\endcsname%
      \put(1364,2506){\makebox(0,0)[l]{\strut{}$\widehat{\sigma}=0.8$}}%
      \csname LTb\endcsname%
      \put(1364,2286){\makebox(0,0)[l]{\strut{}$\widehat{\sigma}=1.0$}}%
    }%
    \gplbacktext
    \put(0,0){\includegraphics{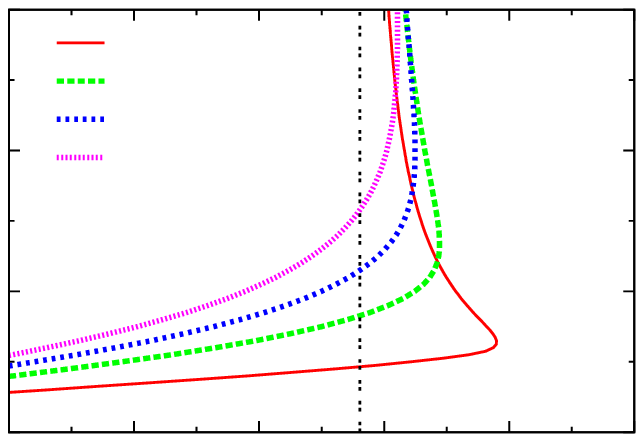}}%
    \gplfronttext
  \end{picture}%
\endgroup

%% file: phase_sigma006.tex
\begingroup
  \makeatletter
  \providecommand\color[2][]{%
    \GenericError{(gnuplot) \space\space\space\@spaces}{%
      Package color not loaded in conjunction with
      terminal option `colourtext'%
    }{See the gnuplot documentation for explanation.%
    }{Either use 'blacktext' in gnuplot or load the package
      color.sty in LaTeX.}%
    \renewcommand\color[2][]{}%
  }%
  \providecommand\includegraphics[2][]{%
    \GenericError{(gnuplot) \space\space\space\@spaces}{%
      Package graphicx or graphics not loaded%
    }{See the gnuplot documentation for explanation.%
    }{The gnuplot epslatex terminal needs graphicx.sty or graphics.sty.}%
    \renewcommand\includegraphics[2][]{}%
  }%
  \providecommand\rotatebox[2]{#2}%
  \@ifundefined{ifGPcolor}{%
    \newif\ifGPcolor
    \GPcolortrue
  }{}%
  \@ifundefined{ifGPblacktext}{%
    \newif\ifGPblacktext
    \GPblacktexttrue
  }{}%
  \let\gplgaddtomacro\g@addto@macro
  \gdef\gplbacktext{}%
  \gdef\gplfronttext{}%
  \makeatother
  \ifGPblacktext
    \def\colorrgb#1{}%
    \def\colorgray#1{}%
  \else
    \ifGPcolor
      \def\colorrgb#1{\color[rgb]{#1}}%
      \def\colorgray#1{\color[gray]{#1}}%
      \expandafter\def\csname LTw\endcsname{\color{white}}%
      \expandafter\def\csname LTb\endcsname{\color{black}}%
      \expandafter\def\csname LTa\endcsname{\color{black}}%
      \expandafter\def\csname LT0\endcsname{\color[rgb]{1,0,0}}%
      \expandafter\def\csname LT1\endcsname{\color[rgb]{0,1,0}}%
      \expandafter\def\csname LT2\endcsname{\color[rgb]{0,0,1}}%
      \expandafter\def\csname LT3\endcsname{\color[rgb]{1,0,1}}%
      \expandafter\def\csname LT4\endcsname{\color[rgb]{0,1,1}}%
      \expandafter\def\csname LT5\endcsname{\color[rgb]{1,1,0}}%
      \expandafter\def\csname LT6\endcsname{\color[rgb]{0,0,0}}%
      \expandafter\def\csname LT7\endcsname{\color[rgb]{1,0.3,0}}%
      \expandafter\def\csname LT8\endcsname{\color[rgb]{0.5,0.5,0.5}}%
    \else
      \def\colorrgb#1{\color{black}}%
      \def\colorgray#1{\color[gray]{#1}}%
      \expandafter\def\csname LTw\endcsname{\color{white}}%
      \expandafter\def\csname LTb\endcsname{\color{black}}%
      \expandafter\def\csname LTa\endcsname{\color{black}}%
      \expandafter\def\csname LT0\endcsname{\color{black}}%
      \expandafter\def\csname LT1\endcsname{\color{black}}%
      \expandafter\def\csname LT2\endcsname{\color{black}}%
      \expandafter\def\csname LT3\endcsname{\color{black}}%
      \expandafter\def\csname LT4\endcsname{\color{black}}%
      \expandafter\def\csname LT5\endcsname{\color{black}}%
      \expandafter\def\csname LT6\endcsname{\color{black}}%
      \expandafter\def\csname LT7\endcsname{\color{black}}%
      \expandafter\def\csname LT8\endcsname{\color{black}}%
    \fi
  \fi
  \setlength{\unitlength}{0.0500bp}%
  \begin{picture}(4680.00,3402.00)%
    \gplgaddtomacro\gplbacktext{%
      \csname LTb\endcsname%
      \put(726,704){\makebox(0,0)[r]{\strut{}$0.42$}}%
      \put(726,1515){\makebox(0,0)[r]{\strut{}$0.47$}}%
      \put(726,2326){\makebox(0,0)[r]{\strut{}$0.52$}}%
      \put(726,3137){\makebox(0,0)[r]{\strut{}$0.57$}}%
      \put(858,484){\makebox(0,0){\strut{}$5$}}%
      \put(1714,484){\makebox(0,0){\strut{}$6$}}%
      \put(2571,484){\makebox(0,0){\strut{}$7$}}%
      \put(3427,484){\makebox(0,0){\strut{}$8$}}%
      \put(4283,484){\makebox(0,0){\strut{}$9$}}%
      \put(220,1920){\rotatebox{-270}{\makebox(0,0){\strut{}$\wh T$}}}%
      \put(2570,154){\makebox(0,0){\strut{}$\widehat{\varphi}$}}%
      \put(1500,2002){\makebox(0,0){\scriptsize FLUID}}%
      \put(2656,2569){\makebox(0,0){\scriptsize REPULSIVE}}%
      \put(2656,2407){\makebox(0,0){\scriptsize GLASS}}%
      \put(3084,1353){\makebox(0,0){\scriptsize ATTRACTIVE}}%
      \put(3084,1191){\makebox(0,0){\scriptsize GLASS}}%
      \put(2142,1839){\makebox(0,0)[l]{\strut{}${\mathsf C}$}}%
      \put(3855,2813){\makebox(0,0)[l]{\strut{}${\mathsf A}_3$}}%
      \put(1860,1191){\makebox(0,0)[l]{\strut{}${\mathsf A}'_3$}}%
      \put(986,2853){\makebox(0,0)[l]{\strut{}$\widehat{\sigma} = 0.06$}}%
    }%
    \gplgaddtomacro\gplfronttext{%
    }%
    \gplbacktext
    \put(0,0){\includegraphics{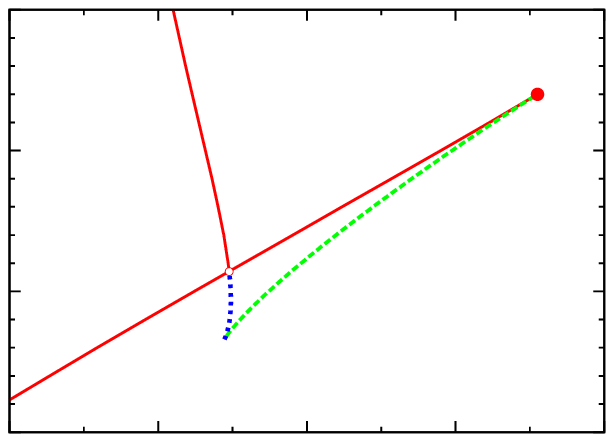}}%
    \gplfronttext
  \end{picture}%
\endgroup

%% file: FV_T047sigma006.tex
\begingroup
  \makeatletter
  \providecommand\color[2][]{%
    \GenericError{(gnuplot) \space\space\space\@spaces}{%
      Package color not loaded in conjunction with
      terminal option `colourtext'%
    }{See the gnuplot documentation for explanation.%
    }{Either use 'blacktext' in gnuplot or load the package
      color.sty in LaTeX.}%
    \renewcommand\color[2][]{}%
  }%
  \providecommand\includegraphics[2][]{%
    \GenericError{(gnuplot) \space\space\space\@spaces}{%
      Package graphicx or graphics not loaded%
    }{See the gnuplot documentation for explanation.%
    }{The gnuplot epslatex terminal needs graphicx.sty or graphics.sty.}%
    \renewcommand\includegraphics[2][]{}%
  }%
  \providecommand\rotatebox[2]{#2}%
  \@ifundefined{ifGPcolor}{%
    \newif\ifGPcolor
    \GPcolortrue
  }{}%
  \@ifundefined{ifGPblacktext}{%
    \newif\ifGPblacktext
    \GPblacktexttrue
  }{}%
  \let\gplgaddtomacro\g@addto@macro
  \gdef\gplbacktext{}%
  \gdef\gplfronttext{}%
  \makeatother
  \ifGPblacktext
    \def\colorrgb#1{}%
    \def\colorgray#1{}%
  \else
    \ifGPcolor
      \def\colorrgb#1{\color[rgb]{#1}}%
      \def\colorgray#1{\color[gray]{#1}}%
      \expandafter\def\csname LTw\endcsname{\color{white}}%
      \expandafter\def\csname LTb\endcsname{\color{black}}%
      \expandafter\def\csname LTa\endcsname{\color{black}}%
      \expandafter\def\csname LT0\endcsname{\color[rgb]{1,0,0}}%
      \expandafter\def\csname LT1\endcsname{\color[rgb]{0,1,0}}%
      \expandafter\def\csname LT2\endcsname{\color[rgb]{0,0,1}}%
      \expandafter\def\csname LT3\endcsname{\color[rgb]{1,0,1}}%
      \expandafter\def\csname LT4\endcsname{\color[rgb]{0,1,1}}%
      \expandafter\def\csname LT5\endcsname{\color[rgb]{1,1,0}}%
      \expandafter\def\csname LT6\endcsname{\color[rgb]{0,0,0}}%
      \expandafter\def\csname LT7\endcsname{\color[rgb]{1,0.3,0}}%
      \expandafter\def\csname LT8\endcsname{\color[rgb]{0.5,0.5,0.5}}%
    \else
      \def\colorrgb#1{\color{black}}%
      \def\colorgray#1{\color[gray]{#1}}%
      \expandafter\def\csname LTw\endcsname{\color{white}}%
      \expandafter\def\csname LTb\endcsname{\color{black}}%
      \expandafter\def\csname LTa\endcsname{\color{black}}%
      \expandafter\def\csname LT0\endcsname{\color{black}}%
      \expandafter\def\csname LT1\endcsname{\color{black}}%
      \expandafter\def\csname LT2\endcsname{\color{black}}%
      \expandafter\def\csname LT3\endcsname{\color{black}}%
      \expandafter\def\csname LT4\endcsname{\color{black}}%
      \expandafter\def\csname LT5\endcsname{\color{black}}%
      \expandafter\def\csname LT6\endcsname{\color{black}}%
      \expandafter\def\csname LT7\endcsname{\color{black}}%
      \expandafter\def\csname LT8\endcsname{\color{black}}%
    \fi
  \fi
  \setlength{\unitlength}{0.0500bp}%
  \begin{picture}(4680.00,3402.00)%
    \gplgaddtomacro\gplbacktext{%
      \csname LTb\endcsname%
      \put(682,704){\makebox(0,0)[r]{\strut{}$2.5$}}%
      \put(682,1515){\makebox(0,0)[r]{\strut{}$2.7$}}%
      \put(682,2326){\makebox(0,0)[r]{\strut{}$2.9$}}%
      \put(682,3137){\makebox(0,0)[r]{\strut{}$3.1$}}%
      \put(814,484){\makebox(0,0){\strut{}$10^{-4}$}}%
      \put(1621,484){\makebox(0,0){\strut{}$10^{-3}$}}%
      \put(2427,484){\makebox(0,0){\strut{}$10^{-2}$}}%
      \put(3234,484){\makebox(0,0){\strut{}$10^{-1}$}}%
      \put(4040,484){\makebox(0,0){\strut{}$10^{0}$}}%
      \put(176,1920){\rotatebox{-270}{\makebox(0,0){\strut{}${\cal V}_{\scriptscriptstyle \rm FP}$}}}%
      \put(2548,154){\makebox(0,0){\strut{}$\widehat{A}$}}%
      \put(2106,947){\makebox(0,0)[l]{\strut{}$\wh{T}=0.47, \ \wh\s = 0.06$}}%
    }%
    \gplgaddtomacro\gplfronttext{%
      \csname LTb\endcsname%
      \put(3566,2926){\makebox(0,0)[r]{\strut{}${\cal F}_{\scriptscriptstyle 1}(\wh{A})$}}%
    }%
    \gplbacktext
    \put(0,0){\includegraphics{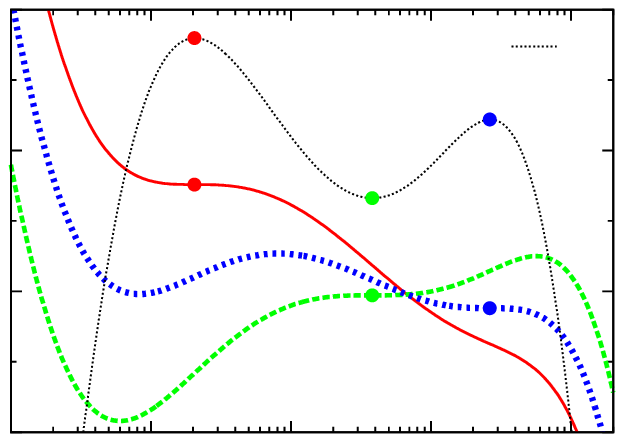}}%
    \gplfronttext
  \end{picture}%
\endgroup

%% file: OP_sigma006T047.tex
\begingroup
  \makeatletter
  \providecommand\color[2][]{%
    \GenericError{(gnuplot) \space\space\space\@spaces}{%
      Package color not loaded in conjunction with
      terminal option `colourtext'%
    }{See the gnuplot documentation for explanation.%
    }{Either use 'blacktext' in gnuplot or load the package
      color.sty in LaTeX.}%
    \renewcommand\color[2][]{}%
  }%
  \providecommand\includegraphics[2][]{%
    \GenericError{(gnuplot) \space\space\space\@spaces}{%
      Package graphicx or graphics not loaded%
    }{See the gnuplot documentation for explanation.%
    }{The gnuplot epslatex terminal needs graphicx.sty or graphics.sty.}%
    \renewcommand\includegraphics[2][]{}%
  }%
  \providecommand\rotatebox[2]{#2}%
  \@ifundefined{ifGPcolor}{%
    \newif\ifGPcolor
    \GPcolortrue
  }{}%
  \@ifundefined{ifGPblacktext}{%
    \newif\ifGPblacktext
    \GPblacktexttrue
  }{}%
  \let\gplgaddtomacro\g@addto@macro
  \gdef\gplbacktext{}%
  \gdef\gplfronttext{}%
  \makeatother
  \ifGPblacktext
    \def\colorrgb#1{}%
    \def\colorgray#1{}%
  \else
    \ifGPcolor
      \def\colorrgb#1{\color[rgb]{#1}}%
      \def\colorgray#1{\color[gray]{#1}}%
      \expandafter\def\csname LTw\endcsname{\color{white}}%
      \expandafter\def\csname LTb\endcsname{\color{black}}%
      \expandafter\def\csname LTa\endcsname{\color{black}}%
      \expandafter\def\csname LT0\endcsname{\color[rgb]{1,0,0}}%
      \expandafter\def\csname LT1\endcsname{\color[rgb]{0,1,0}}%
      \expandafter\def\csname LT2\endcsname{\color[rgb]{0,0,1}}%
      \expandafter\def\csname LT3\endcsname{\color[rgb]{1,0,1}}%
      \expandafter\def\csname LT4\endcsname{\color[rgb]{0,1,1}}%
      \expandafter\def\csname LT5\endcsname{\color[rgb]{1,1,0}}%
      \expandafter\def\csname LT6\endcsname{\color[rgb]{0,0,0}}%
      \expandafter\def\csname LT7\endcsname{\color[rgb]{1,0.3,0}}%
      \expandafter\def\csname LT8\endcsname{\color[rgb]{0.5,0.5,0.5}}%
    \else
      \def\colorrgb#1{\color{black}}%
      \def\colorgray#1{\color[gray]{#1}}%
      \expandafter\def\csname LTw\endcsname{\color{white}}%
      \expandafter\def\csname LTb\endcsname{\color{black}}%
      \expandafter\def\csname LTa\endcsname{\color{black}}%
      \expandafter\def\csname LT0\endcsname{\color{black}}%
      \expandafter\def\csname LT1\endcsname{\color{black}}%
      \expandafter\def\csname LT2\endcsname{\color{black}}%
      \expandafter\def\csname LT3\endcsname{\color{black}}%
      \expandafter\def\csname LT4\endcsname{\color{black}}%
      \expandafter\def\csname LT5\endcsname{\color{black}}%
      \expandafter\def\csname LT6\endcsname{\color{black}}%
      \expandafter\def\csname LT7\endcsname{\color{black}}%
      \expandafter\def\csname LT8\endcsname{\color{black}}%
    \fi
  \fi
  \setlength{\unitlength}{0.0500bp}%
  \begin{picture}(4680.00,3402.00)%
    \gplgaddtomacro\gplbacktext{%
      \csname LTb\endcsname%
      \put(682,704){\makebox(0,0)[r]{\strut{}$0$}}%
      \put(682,1312){\makebox(0,0)[r]{\strut{}$0.5$}}%
      \put(682,1921){\makebox(0,0)[r]{\strut{}$1$}}%
      \put(682,2529){\makebox(0,0)[r]{\strut{}$1.5$}}%
      \put(682,3137){\makebox(0,0)[r]{\strut{}$2$}}%
      \put(814,484){\makebox(0,0){\strut{}$0.76$}}%
      \put(1681,484){\makebox(0,0){\strut{}$0.81$}}%
      \put(2549,484){\makebox(0,0){\strut{}$0.86$}}%
      \put(3416,484){\makebox(0,0){\strut{}$0.91$}}%
      \put(4283,484){\makebox(0,0){\strut{}$0.96$}}%
      \put(176,1920){\rotatebox{-270}{\makebox(0,0){\strut{}$-\log \sqrt{\widehat{A}}$}}}%
      \put(2548,154){\makebox(0,0){\strut{}$\log \widehat{\varphi}$}}%
      \put(987,2894){\makebox(0,0)[l]{\strut{}$\widehat{\sigma} = 0.06$}}%
    }%
    \gplgaddtomacro\gplfronttext{%
      \csname LTb\endcsname%
      \put(3745,1169){\makebox(0,0)[r]{\strut{}$\widehat{T}=0.47$}}%
      \csname LTb\endcsname%
      \put(3745,883){\makebox(0,0)[r]{\strut{}$\widehat{T}=0.50$}}%
    }%
    \gplbacktext
    \put(0,0){\includegraphics{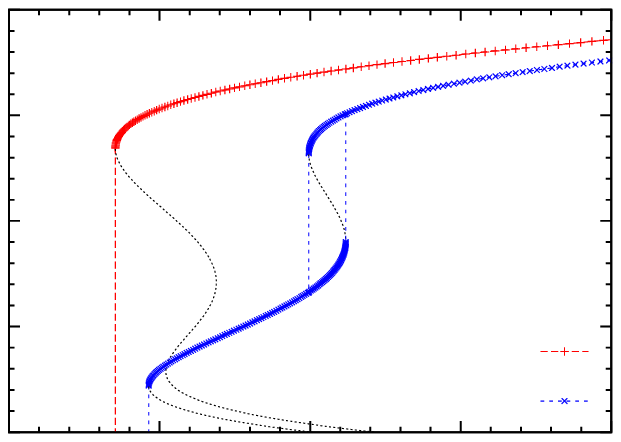}}%
    \gplfronttext
  \end{picture}%
\endgroup

%% file: phase_global.tex
\begingroup
  \makeatletter
  \providecommand\color[2][]{%
    \GenericError{(gnuplot) \space\space\space\@spaces}{%
      Package color not loaded in conjunction with
      terminal option `colourtext'%
    }{See the gnuplot documentation for explanation.%
    }{Either use 'blacktext' in gnuplot or load the package
      color.sty in LaTeX.}%
    \renewcommand\color[2][]{}%
  }%
  \providecommand\includegraphics[2][]{%
    \GenericError{(gnuplot) \space\space\space\@spaces}{%
      Package graphicx or graphics not loaded%
    }{See the gnuplot documentation for explanation.%
    }{The gnuplot epslatex terminal needs graphicx.sty or graphics.sty.}%
    \renewcommand\includegraphics[2][]{}%
  }%
  \providecommand\rotatebox[2]{#2}%
  \@ifundefined{ifGPcolor}{%
    \newif\ifGPcolor
    \GPcolortrue
  }{}%
  \@ifundefined{ifGPblacktext}{%
    \newif\ifGPblacktext
    \GPblacktexttrue
  }{}%
  \let\gplgaddtomacro\g@addto@macro
  \gdef\gplbacktext{}%
  \gdef\gplfronttext{}%
  \makeatother
  \ifGPblacktext
    \def\colorrgb#1{}%
    \def\colorgray#1{}%
  \else
    \ifGPcolor
      \def\colorrgb#1{\color[rgb]{#1}}%
      \def\colorgray#1{\color[gray]{#1}}%
      \expandafter\def\csname LTw\endcsname{\color{white}}%
      \expandafter\def\csname LTb\endcsname{\color{black}}%
      \expandafter\def\csname LTa\endcsname{\color{black}}%
      \expandafter\def\csname LT0\endcsname{\color[rgb]{1,0,0}}%
      \expandafter\def\csname LT1\endcsname{\color[rgb]{0,1,0}}%
      \expandafter\def\csname LT2\endcsname{\color[rgb]{0,0,1}}%
      \expandafter\def\csname LT3\endcsname{\color[rgb]{1,0,1}}%
      \expandafter\def\csname LT4\endcsname{\color[rgb]{0,1,1}}%
      \expandafter\def\csname LT5\endcsname{\color[rgb]{1,1,0}}%
      \expandafter\def\csname LT6\endcsname{\color[rgb]{0,0,0}}%
      \expandafter\def\csname LT7\endcsname{\color[rgb]{1,0.3,0}}%
      \expandafter\def\csname LT8\endcsname{\color[rgb]{0.5,0.5,0.5}}%
    \else
      \def\colorrgb#1{\color{black}}%
      \def\colorgray#1{\color[gray]{#1}}%
      \expandafter\def\csname LTw\endcsname{\color{white}}%
      \expandafter\def\csname LTb\endcsname{\color{black}}%
      \expandafter\def\csname LTa\endcsname{\color{black}}%
      \expandafter\def\csname LT0\endcsname{\color{black}}%
      \expandafter\def\csname LT1\endcsname{\color{black}}%
      \expandafter\def\csname LT2\endcsname{\color{black}}%
      \expandafter\def\csname LT3\endcsname{\color{black}}%
      \expandafter\def\csname LT4\endcsname{\color{black}}%
      \expandafter\def\csname LT5\endcsname{\color{black}}%
      \expandafter\def\csname LT6\endcsname{\color{black}}%
      \expandafter\def\csname LT7\endcsname{\color{black}}%
      \expandafter\def\csname LT8\endcsname{\color{black}}%
    \fi
  \fi
  \setlength{\unitlength}{0.0500bp}%
  \begin{picture}(4680.00,3402.00)%
    \gplgaddtomacro\gplbacktext{%
      \csname LTb\endcsname%
      \put(682,704){\makebox(0,0)[r]{\strut{}$0$}}%
      \put(682,1515){\makebox(0,0)[r]{\strut{}$0.7$}}%
      \put(682,2326){\makebox(0,0)[r]{\strut{}$1.4$}}%
      \put(682,3137){\makebox(0,0)[r]{\strut{}$2.1$}}%
      \put(814,484){\makebox(0,0){\strut{}$3$}}%
      \put(1970,484){\makebox(0,0){\strut{}$5$}}%
      \put(3127,484){\makebox(0,0){\strut{}$7$}}%
      \put(4283,484){\makebox(0,0){\strut{}$9$}}%
      \put(176,1920){\rotatebox{-270}{\makebox(0,0){\strut{}$-\b U_0 -\ln \wh \s$}}}%
      \put(2548,154){\makebox(0,0){\strut{}$\widehat{\varphi}$}}%
      \put(1392,1921){\makebox(0,0){\scriptsize FLUID}}%
      \put(3271,2905){\makebox(0,0){\scriptsize REPULSIVE}}%
      \put(3271,2732){\makebox(0,0){\scriptsize GLASS}}%
      \put(3416,1110){\makebox(0,0){\scriptsize ATTRACTIVE}}%
      \put(3416,936){\makebox(0,0){\scriptsize GLASS}}%
      \put(2693,2152){\makebox(0,0)[l]{\strut{}${\mathsf C}$}}%
      \put(2462,1052){\makebox(0,0)[l]{\strut{}${\mathsf A}_4$}}%
      \put(3994,1631){\makebox(0,0)[l]{\strut{}${\mathsf A}_3$}}%
    }%
    \gplgaddtomacro\gplfronttext{%
      \csname LTb\endcsname%
      \put(1826,2969){\makebox(0,0)[r]{\strut{}$\widehat{\sigma}=.06$}}%
      \csname LTb\endcsname%
      \put(1826,2749){\makebox(0,0)[r]{\strut{}$\widehat{\sigma}=.12$}}%
      \csname LTb\endcsname%
      \put(1826,2529){\makebox(0,0)[r]{\strut{}$\widehat{\sigma}=.19$}}%
    }%
    \gplbacktext
    \put(0,0){\includegraphics{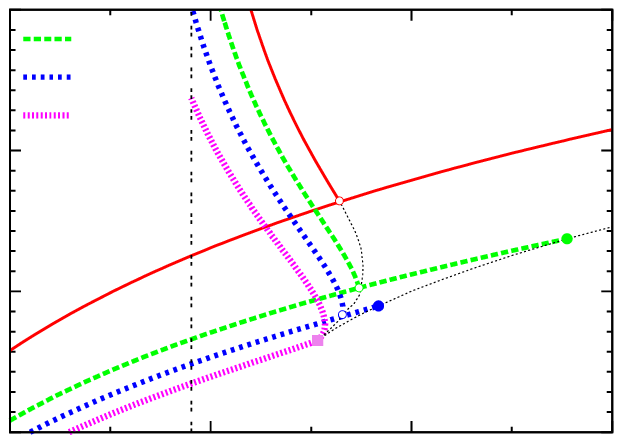}}%
    \gplfronttext
  \end{picture}%
\endgroup

%% file: phasediagram_ss.tex
\begingroup
  \makeatletter
  \providecommand\color[2][]{%
    \GenericError{(gnuplot) \space\space\space\@spaces}{%
      Package color not loaded in conjunction with
      terminal option `colourtext'%
    }{See the gnuplot documentation for explanation.%
    }{Either use 'blacktext' in gnuplot or load the package
      color.sty in LaTeX.}%
    \renewcommand\color[2][]{}%
  }%
  \providecommand\includegraphics[2][]{%
    \GenericError{(gnuplot) \space\space\space\@spaces}{%
      Package graphicx or graphics not loaded%
    }{See the gnuplot documentation for explanation.%
    }{The gnuplot epslatex terminal needs graphicx.sty or graphics.sty.}%
    \renewcommand\includegraphics[2][]{}%
  }%
  \providecommand\rotatebox[2]{#2}%
  \@ifundefined{ifGPcolor}{%
    \newif\ifGPcolor
    \GPcolortrue
  }{}%
  \@ifundefined{ifGPblacktext}{%
    \newif\ifGPblacktext
    \GPblacktexttrue
  }{}%
  \let\gplgaddtomacro\g@addto@macro
  \gdef\gplbacktext{}%
  \gdef\gplfronttext{}%
  \makeatother
  \ifGPblacktext
    \def\colorrgb#1{}%
    \def\colorgray#1{}%
  \else
    \ifGPcolor
      \def\colorrgb#1{\color[rgb]{#1}}%
      \def\colorgray#1{\color[gray]{#1}}%
      \expandafter\def\csname LTw\endcsname{\color{white}}%
      \expandafter\def\csname LTb\endcsname{\color{black}}%
      \expandafter\def\csname LTa\endcsname{\color{black}}%
      \expandafter\def\csname LT0\endcsname{\color[rgb]{1,0,0}}%
      \expandafter\def\csname LT1\endcsname{\color[rgb]{0,1,0}}%
      \expandafter\def\csname LT2\endcsname{\color[rgb]{0,0,1}}%
      \expandafter\def\csname LT3\endcsname{\color[rgb]{1,0,1}}%
      \expandafter\def\csname LT4\endcsname{\color[rgb]{0,1,1}}%
      \expandafter\def\csname LT5\endcsname{\color[rgb]{1,1,0}}%
      \expandafter\def\csname LT6\endcsname{\color[rgb]{0,0,0}}%
      \expandafter\def\csname LT7\endcsname{\color[rgb]{1,0.3,0}}%
      \expandafter\def\csname LT8\endcsname{\color[rgb]{0.5,0.5,0.5}}%
    \else
      \def\colorrgb#1{\color{black}}%
      \def\colorgray#1{\color[gray]{#1}}%
      \expandafter\def\csname LTw\endcsname{\color{white}}%
      \expandafter\def\csname LTb\endcsname{\color{black}}%
      \expandafter\def\csname LTa\endcsname{\color{black}}%
      \expandafter\def\csname LT0\endcsname{\color{black}}%
      \expandafter\def\csname LT1\endcsname{\color{black}}%
      \expandafter\def\csname LT2\endcsname{\color{black}}%
      \expandafter\def\csname LT3\endcsname{\color{black}}%
      \expandafter\def\csname LT4\endcsname{\color{black}}%
      \expandafter\def\csname LT5\endcsname{\color{black}}%
      \expandafter\def\csname LT6\endcsname{\color{black}}%
      \expandafter\def\csname LT7\endcsname{\color{black}}%
      \expandafter\def\csname LT8\endcsname{\color{black}}%
    \fi
  \fi
  \setlength{\unitlength}{0.0500bp}%
  \begin{picture}(4680.00,3402.00)%
    \gplgaddtomacro\gplbacktext{%
      \csname LTb\endcsname%
      \put(550,704){\makebox(0,0)[r]{\strut{}$0$}}%
      \put(550,1515){\makebox(0,0)[r]{\strut{}$2$}}%
      \put(550,2326){\makebox(0,0)[r]{\strut{}$4$}}%
      \put(550,3137){\makebox(0,0)[r]{\strut{}$6$}}%
      \put(682,484){\makebox(0,0){\strut{}$2$}}%
      \put(1582,484){\makebox(0,0){\strut{}$3$}}%
      \put(2483,484){\makebox(0,0){\strut{}$4$}}%
      \put(3383,484){\makebox(0,0){\strut{}$5$}}%
      \put(4283,484){\makebox(0,0){\strut{}$6$}}%
      \put(176,1920){\rotatebox{-270}{\makebox(0,0){\strut{}$\wh{T}$}}}%
      \put(2482,154){\makebox(0,0){\strut{}$\widehat{\varphi}$}}%
      \put(1582,1718){\makebox(0,0){\scriptsize FLUID}}%
      \put(3743,1718){\makebox(0,0){\scriptsize GLASS}}%
    }%
    \gplgaddtomacro\gplfronttext{%
      \csname LTb\endcsname%
      \put(1400,2946){\makebox(0,0)[l]{\strut{}$\widehat{\sigma}=0.8$}}%
      \csname LTb\endcsname%
      \put(1400,2726){\makebox(0,0)[l]{\strut{}$\widehat{\sigma}=0.4$}}%
      \csname LTb\endcsname%
      \put(1400,2506){\makebox(0,0)[l]{\strut{}$\widehat{\sigma}=0.2$}}%
      \csname LTb\endcsname%
      \put(1400,2286){\makebox(0,0)[l]{\strut{}$\widehat{\sigma}=0.1$}}%
    }%
    \gplbacktext
    \put(0,0){\includegraphics{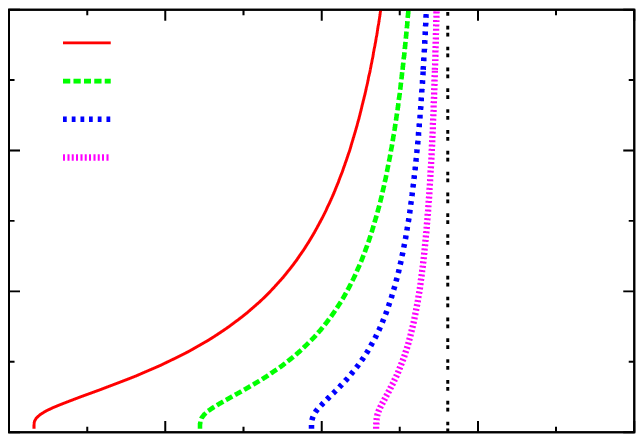}}%
    \gplfronttext
  \end{picture}%
\endgroup